\documentclass[twocolumn,showpacs,preprintnumbers,prc,amsmath,amssymb,floatfix]{revtex4}

\usepackage{graphicx}
\usepackage{dcolumn}
\usepackage{bm}

\newcommand{\beqa}{\begin{equation}}
\newcommand{\eeqa}{\end{equation}}

\begin{document}

\title{Investigating photoproduction of scalar mesons at medium energies}

\author{M. L. L. da Silva$^1$ and M. V. T. Machado$^2$}
\affiliation{$^1$ Instituto de F\'{\i}sica e Matem\'atica, Universidade Federal de
Pelotas\\
Caixa Postal 354, CEP 96010-090, Pelotas, RS, Brazil\\
$^2$ High Energy Physics Phenomenology Group, GFPAE  IF-UFRGS \\
Caixa Postal 15051, CEP 91501-970, Porto Alegre, RS, Brazil}

\begin{abstract}
In this letter we study the photoproduction of scalar mesons in the intermediate energies considering distinct
mixing scenarios in the description of meson physical states. The differential and integrated total cross
section are computed for the cases of the mesons $a_0(980)$, $f_0(1500)$ and $f_0(1710)$ focusing the GlueX
energy regime with photon energy $E_{\gamma} = 9$ GeV. Our results indicate that light-quark scalar meson
photoproduction is well suited for studying hybrid mesons structure.
\end{abstract}


\maketitle

\section{Introduction}

For a long time, the understanding  of the scalar sector of mesons has been problematic and still a subject of
debate. The low energy scalar states, for instance the $f_0(980)$  and $a_0(980)$ ($J^{PC}=0^{++}$) have been
considered in the past conventional quark-antiquark mesons \cite{Marco1}, tetraquarks \cite{Marco2}, hadron
molecules \cite{Marco3}, glueballs and hybrids \cite{Marco4,Ochs}. In addition, for scalar mesons $f_0(1500)$
and $f_0(1710)$ there is yet no consensus on their status \cite{Lee,Kirk}. Such a confuse interpretation comes
from the fact that despite the QCD providing a clear description of the strong interaction of partons
(quarks/gluons) at high energies, the situation is complex in the low energy regime. Namely, obtaining
quantitative predictions from QCD at low energy, like the spectrum of baryons and mesons, remains challenging and
nowadays relies on numerical techniques of Lattice QCD (LQCD). The current understanding of how quarks form
mesons has evolved within QCD and it is expected a richer spectrum of mesons that takes into account not only
the quark degrees of freedom but also the gluonic ones. A common example are the glueball resonance with no
quarks which are expected to have  quantum numbers not exotic and cannot be accommodated within quark-antiquark
nonets \cite{Crede}. This glueball states can mix with quark-antiquark states with the same quantum numbers.
We also expect that excitations of the gluonic field from the quark binding could generate the so-called hybrid
mesons, which are a quark-antiquark state plus a one or more gluonic degree of freedom.  Thus, the exotic
mesons provide the ideal laboratory for testing QCD in the confinement regime once they explicitly manifest the
gluonic degrees of freedom \cite{LQCD}.

The mesons $f_0(1500)$ and $f_0(1710)$ are considered good candidates for the scalar glueball \cite{Lee,Kirk}.
However, in this mass region, the glueball state will mix strongly with nearby $q\bar{q}$ states \cite{close1,close2}.
In the lowest order, mixture of the scalar glueball $G$ and quarkonia states $n\bar{n}=(u\bar{u}+ d\bar{d})/\sqrt{2}$
and $s\bar{s}$ can be obtained as in Refs. \cite{Lee,Kirk}. The mixing is write in the following form
$|f_0(M)\rangle=c_1 \,|n\bar{n}\rangle +c_2 \,|s\bar{s} \rangle + c_3 \,|G\rangle$, where the normalization
condition is $\sum_{i=1}^{3} c_{i}^{2}=1$. In the literature the parameters have been adjusted to the observed
resonances $f_0(1370)$, $f_0(1500)$ and $f_0(1710)$ and are obtained from the mass of the glueball, $n\bar{n}$ and
$s\bar{s}$ states \cite{Kirk}. The unknown about the mixing parameters still remains, on this way some proposal
to set the parameters is very important to determine the structure of this resonances.

Despite an active experimental program, data supporting the existence of meson states having exotic quantum
numbers are still sparse \cite{Crede,Klempt}. Concerning the scalar sector, the $f_0(1500)$ and $f_0(1710)$ are
possibilities for the ninth member of the SU(3) flavor nonet \cite{PDG}. It is in general assumed the surplus
of isoscalar scalar in the mass region 1300-1700 MeV is due to the presence of a scalar glueball, which was
supported by calculation on quenched LQCD \cite{quenchedLQCD}. Those meson were the viewed as mixed
quark-antiquark and gluonium states. The interpretation has changed when considering unquenched LQCD calculations
\cite{unquenchedLQCD}. As an alternative the radiative transitions could offer a means of probing the structure
of hadrons as the coupling to the charges and spins of the constituents reveals detailed information about the
wavefunctions and could discriminate among models. Radiative decays of $f_0$ mesons to $\rho$ and $\omega$ have
been shown to provide effective probes of their structure. If the $f_0$ mesons are mixed states their radiative
decays to a vector mesons are strongly affected by the degree of mixing between the $q\bar{q}$ state and the
glueball \cite{CDK}. Following \cite{CDK,DK}, assuming that the $q\bar{q}$ contribution to the referred mesons is
in the $1^{3}P_0$ nonet, the discrimination among different mixing possibilities is strong.

In the context referred above, the photoproduction of exotic \cite{DK} is interesting for several reasons. Using
arguments based on vector meson dominance (VDM) the photon can behave like an $S=1$ quark-antiquark system. In
several models, such a system is more likely to couple to exotic quantum number hybrids. Recently, LQCD
calculations have been performed to compute the radiative decay of charmonium and hybrid states \cite{93},
verifying a large radiative decay for an exotic quantum number hybrid. Based on these results in the charmonium
sector, photoproduction appears to be a good place to look for hybrid mesons. In particular, the photoproduction of
scalar mesons at intermediate energies could provide an alternative to the direct observation of the radiative
decays. From the experimental point of view, the GlueX experiment \cite{GlueX} is being installed and it is located
in JLab accelerator. Its primary purpose is to understand the nature of confinement in QCD by mapping the spectrum
of exotic mesons generated buy the excitation of the gluonic field binding the quarks. The experiment will be able
to probe new areas by using photoproduction to produce exotic states.

Here, we will focus on photoproduction of mesons states $a_0(980)$, $f_0(1500)$ and $f_0(1710)$. The $f_0(1500)$
and $f_0(1710)$ mesons will be considered distinct in distinct mixing possibilities and assuming that $a_0(980)$ is
member of the ground-state nonet. This is an established idea although the motivation is different from some other
works in the literature, e.g. Ref. \cite{BB}. The theoretical formalism employed is the Regge approach with reggeized $\rho$
and $\omega$ exchange \cite{DK}. This paper is organized as follows: in next  section we present the main
expressions for scattering amplitudes and cross section calculation of scalar meson photoproduction in the Regge
theory and in last section we shown the numerical results discussing distinct mixing scenarios and main conclusions.

\section{Model and cross section calculation}
The reaction proposed is $\gamma p \rightarrow p\,M$, where $M$ is either of the resonances $a_0(980)$, $f_0(1500)$
or $f_0(1710)$. In practice, the meson $M$  will decay into two mesons. The contribution from vector mesons can be
eliminated by considering only the all-neutral channels, that is the $\pi^0\pi^0$, $\eta^0\eta^0$ and $4\pi^0$
decays of the scalar referred mesons. In the scope of Regge theory the differential cross section in the narrow-width
limit for a meson of mass $m_S$ is given by
\begin{eqnarray}
\frac{d\sigma}{dt}(\gamma p \rightarrow p M)= \frac{|{\cal M}(s,t)|^2}{64\pi\,(s-m_p^2)^2},
\label{dsigma}
\end{eqnarray}
where ${\cal M}$ is the scattering amplitude for the process, $s,\,t$ are usual Mandelstan variables and $m_p$ is the
proton mass. For the exchange of a single vector meson (for instance, $\rho$ or $\omega$):
\begin{eqnarray}
|{\cal M} (s,t)|^2 &=& -\frac{1}{2}{\cal A}^2(s,t)\left[\frac{}{}s(t-t_1)(t-t_2) \right.\nonumber\\
      & & \,\,\,\,\,\,\,\, \left.    +\frac{1}{2}t(t^2 - 2 (m_S^2 +s)t + m_S^4) \right] \nonumber\\
& - &{\cal A}(s,t){\cal B}(s,t)m_ps(t-t_1)(t-t_2) \nonumber\\
& - &\frac{1}{8}{\cal B}^2(s,t)s(4m_p^2-t)(t-t_1)(t-t_2).
\label{msquare}
\end{eqnarray}
where $t_1$ and $t_2$ are the kinematical boundaries
\begin{eqnarray}
\!\!\!\!\!\!t_{1,2} &=& \frac{1}{2s}\left[-(m_p^2-s)^2+m_S^2(m_p^2+s)\right. \nonumber\\
& \pm & \left. (m_p^2-s)\sqrt{(m_p^2-s)^2-2m_s^2(m_p^2+s)+m_S^4}\, \right],
\label{t12}
\end{eqnarray}
and where one uses the standard prescription for Reggeising the Feynman propagators  assuming a linear Regge trajectory
$\alpha_V(t)= \alpha_{V0}+ \alpha^\prime_V t$ for writing down the quantities ${\cal A}(s,t)$ and ${\cal B}(s,t)$:
\begin{eqnarray}
{\cal A}(s,t) & = & g_A\,\left(\frac{s}{s_0}\right)^{\alpha_V(t)-1}\frac{\pi\alpha^\prime_V}{\sin(\pi\alpha_V(t))}
\frac{1-e^{-i\pi\alpha_V(t)}}{2\,\Gamma(\alpha_V(t))},\nonumber \\
{\cal B}(s,t) & = & -\frac{g_B}{g_A}\,{\cal A}(s,t).
\label{abdef}
\end{eqnarray}
The Eq.(\ref{msquare}) is different from that presented in Ref. \cite{DK}, which contains a typographical error. However,
we have verified that the numerical results in Ref. \cite{DK} are correct.

Here, it is assumed non-degenerate $\rho$ and $\omega$ trajectories $\alpha_{V}(t) = \alpha_{V}(0)+\alpha^{\prime}_{V}t$,
with $\alpha_{V}(0)=0.55\,(0.44)$ and $\alpha^{\prime}_{V} = 0.8\,(0.9)$ for $\rho$ ($\omega$). In Eq. (\ref{abdef})
above, one has that $g_A = g_S(g_V+2 m_p g_T)$ and $g_B=2g_Sg_T$. The quantities $g_V$ and $g_T$ are the $VNN$ vector and
tensor couplings, $g_S$ is the $\gamma V N$ coupling. The $\omega N N$ couplings are rather well  defined\cite{Holinde},
and we have used $g_V^{\omega} = 15$ and $g_T^\omega =0$ following Ref.\cite{DK}. The $\rho N N$ couplings are not so
well defined and we have considered $g_V^{\rho} = 3.4$, $g_T^{\rho}= 11$ GeV$^{-1}$. The $S V \gamma$ coupling, $g_S$, can
be obtained from the radiative decay width through \cite{KKNHH}
\begin{eqnarray}
\Gamma(S \to \gamma V) = g_S^2\frac{m_S^3}{32\pi}\Bigg(1-\frac{m_V^2}{m_S^2}
\Bigg)^3.
\label{width}
\end{eqnarray}

In the case of the $f_0$ mesons being considered as mixed $n\bar{n}$, $s\bar{s}$ and glueball states their
radiative decays to a vector meson, $S\rightarrow V\gamma$, are expected to be highly sensitive to the degree of
mixing between the quark-antiquark basis and the glueball \cite{CDK}. Here, we will consider three distinct mixing
scenarios: (I) the bare glueball is lighter than the bare $n\bar{n}$ state; (II) the glueball mass is between the
$n\bar{n}$ and $s\bar{s}$ bare state; (III) glueball mass is heavier than the bare $s\bar{s}$ state. For the meson
$a_0(980)$ decay to $\rho$ is assumed that it is a member of the ground-state nonet. The numerical values for the
widths taking into account the effects of mixing on the radiative decays of the scalars on $\rho$ and $\omega$
(in units of keV) are shown in Table \ref{tab1} and for $a_0(980)$ we have
\begin{eqnarray}
 \Gamma(a_0(980)\rightarrow \gamma \rho (\omega)) = 14 (126)\, {\rm keV}\,.
\end{eqnarray}
The widths considered  are taken from Ref. \cite{DK}. Clearly, the width is largely model dependent and other
approaches can be used. We call attention to the interesting work in Ref. \cite{Jacosa}, where the decays  of
a light scalar meson into a vector mesons and a photon ($S\rightarrow V\gamma$) are evaluated in the tetraquark
and quarkonium assignements of the scalar states. The different nature of them corresponds to distinct 
large-$N_c$ dominant interaction Lagrangians.
\begin{table}[t]
\begin{center}
\begin{tabular} {|c|c|c|c|}
\hline
Scenario & $f_0(1500)\rightarrow \gamma V$  & $f_0(1710)\rightarrow \gamma V$  \\
\hline
\hline
(I) & 2519 (280) & 42 (4.7) \\
\hline
(II) & 1458 (162) & 94 (10.4) \\
\hline
(III) & 476 (53) & 705 (78) \\
\hline
\end{tabular}
\end{center}
\caption{\it The widths, $\Gamma(S\rightarrow \gamma V)$, for the radiative decays of the scalar mesons to vector
mesons $V=\rho\,(\omega)$. They are presented in units of keV.}
\label{tab1}
\end{table}

In what follows we present the numerical results for the scalar mesons considered in present study and the consequence
of the different mixing scenarios discussed above.

\section{Results and discussions}

\begin{figure}[ht]
\includegraphics[scale=0.32]{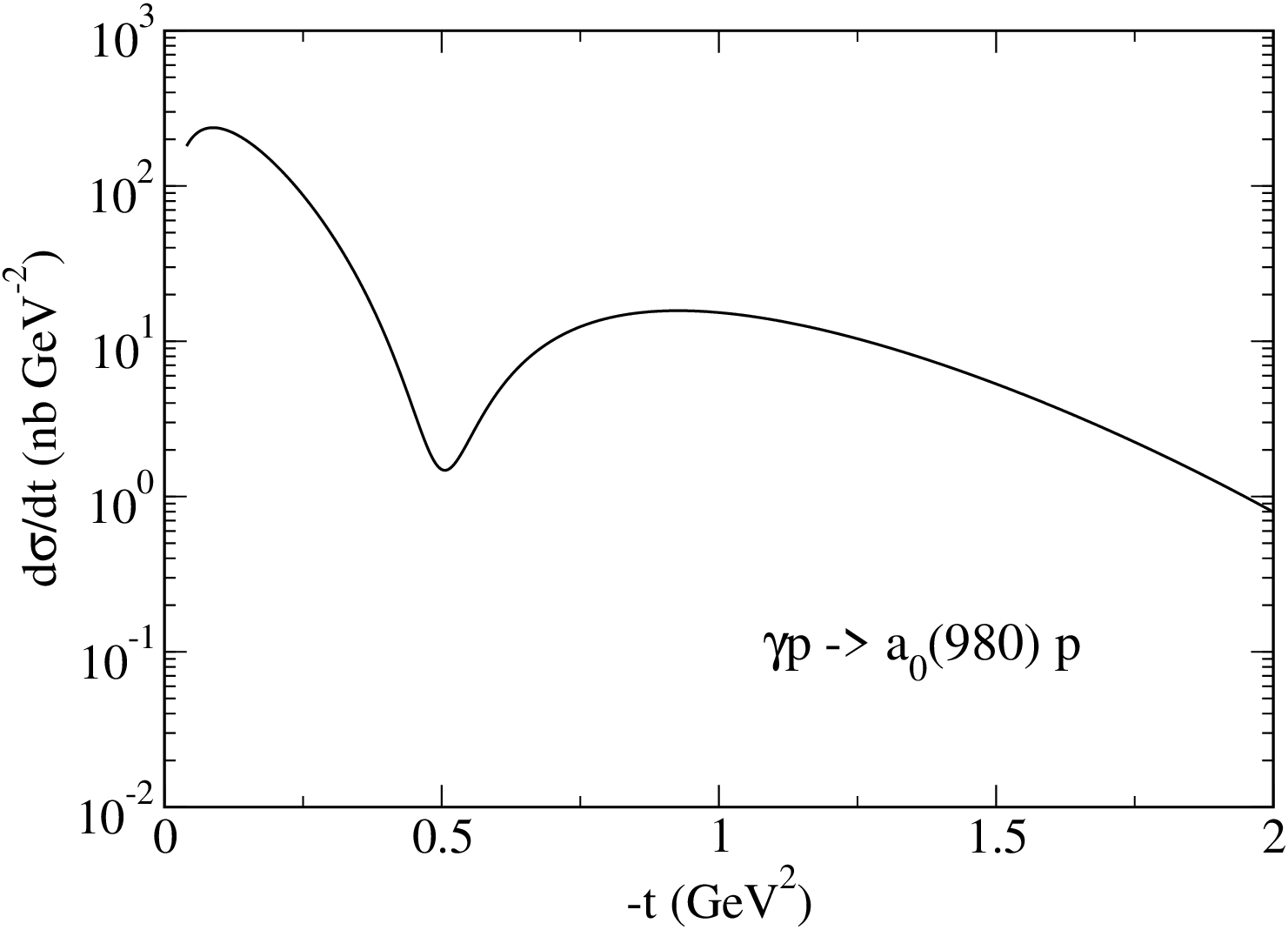}
\caption{(Color online) Differential photoproduction cross section on proton for $a_0(980)$ at GlueX energy $E_{\gamma}=9$ GeV.}
\label{fig:1}
\end{figure}

\begin{figure}[ht]
\includegraphics[scale=0.32]{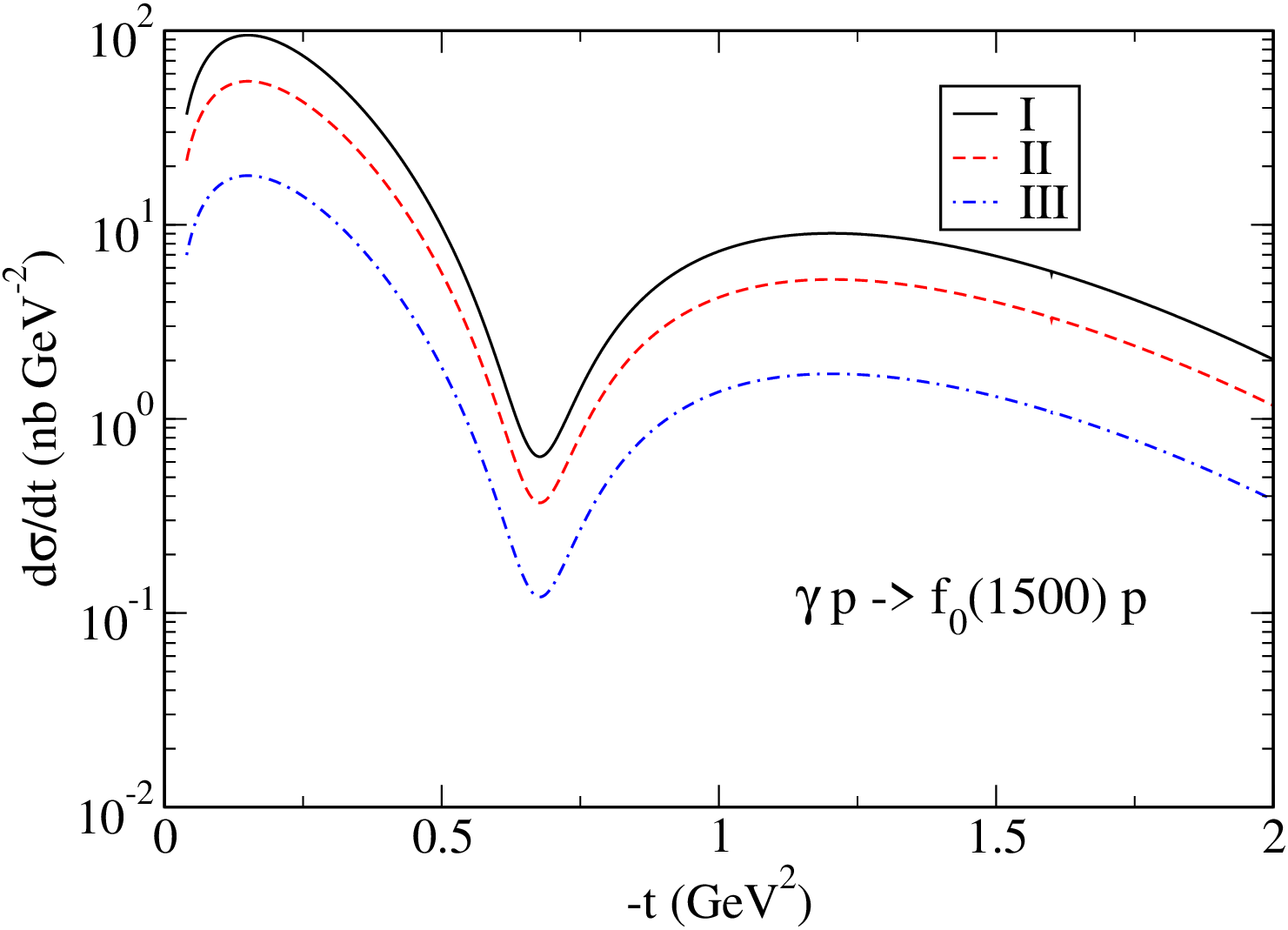}
\caption{(Color online) Differential photoproduction cross section on proton for $f_0(1500)$ at GlueX energy $E_{\gamma}=9$ GeV.
The results for the distinct  three mixing scenarios are presented: I (solid line), II (dashed line) and III
(dot-dashed line).}
\label{fig:2}
\end{figure}

\begin{figure}[ht]
\includegraphics[scale=0.32]{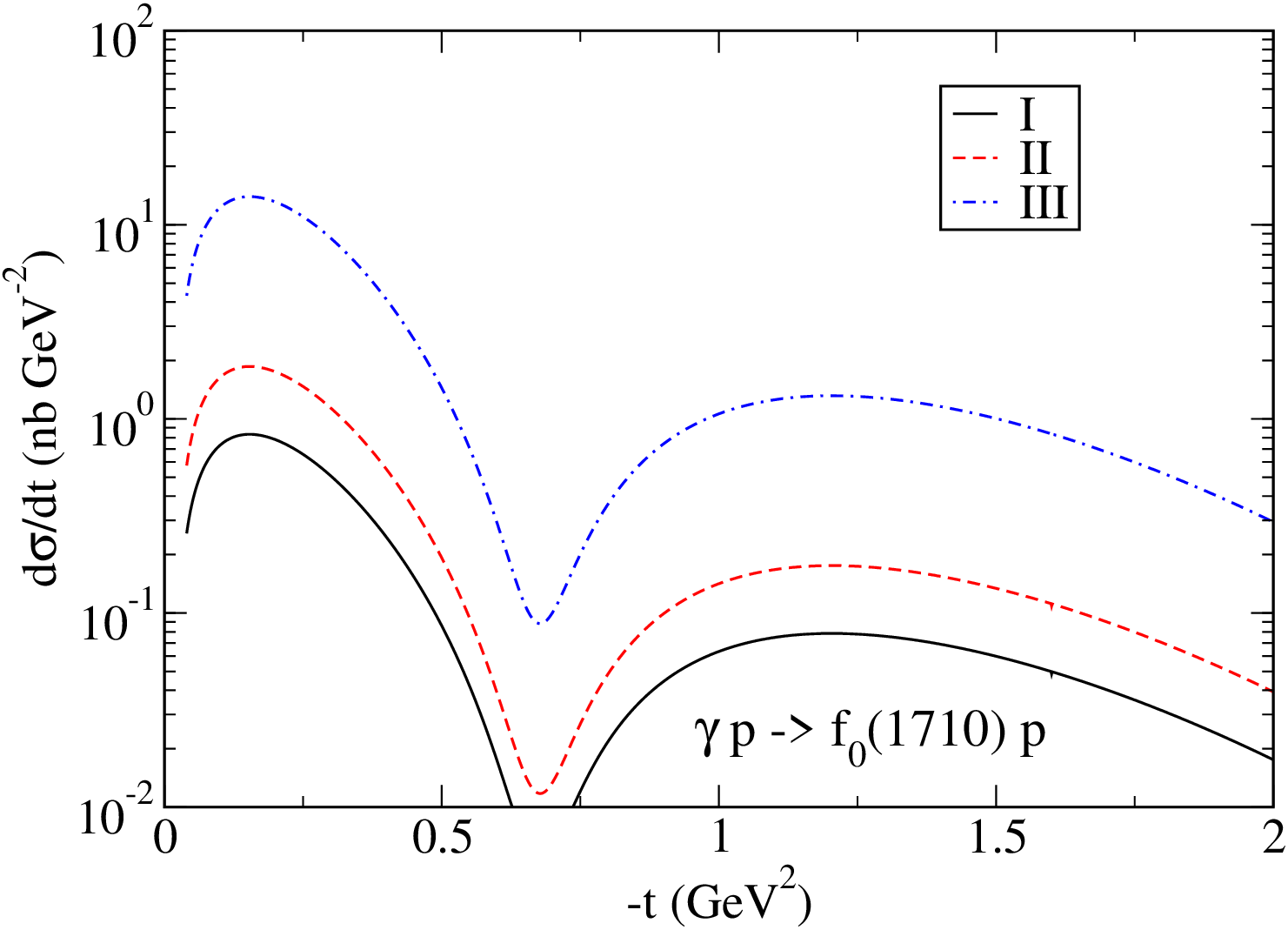}
\caption{(Color online) Differential photoproduction cross section on proton for $f_0(1710)$ at GlueX energy $E_{\gamma}=9$ GeV.
The results for the distinct  three mixing scenarios are presented: I (solid line), II (dashed line) and III
(dot-dashed line).}
\label{fig:3}
\end{figure}

Let us summarize the numerical results for the photoproduction of scalar mesons $a_0(980)$, $f_0(1500)$ and $f_0(1710)$.
The differential cross section for $a_0 (980)$ is presented in Fig. \ref{fig:1} at $E_{\gamma} = 9$ GeV, which is
vanishing in the forward direction due to the helicity flip at the photon-scalar vertex and having a deep dip at
$ -t \approx 0.5$ GeV$^2$. In the current scenario, the forward cross section is sizable, $d\sigma/dt_{t=0}\simeq 1$
nb/GeV$^{-1}$. The differential cross sections for $f_0 (1500)$ are presented in Fig. \ref{fig:2} at $E_{\gamma} = 9$
GeV, and showing the consequences of distinct mixing scenarios. The general structure follows the previous figure. In the
scenario (I) the cross section is higher the other scenarios. That is, a light glueball mass implies a larger cross
section for the $f_0 (1500)$ mesons. On the other hand, the inverse situation occurs for the $f_0(1710)$ mesons as shown
in Fig. \ref{fig:3} where the large cross section comes from the heavy glueball mass component. The cross sections
reflect directly the radiative decay widths as can be verified  from simple inspection of  Table 1. It was advocated in
Ref. \cite{DK} that the ratios of cross sections could  give the ``weigh'' of the glueball content. For completeness, the
integrated cross sections for photoproduction of the scalars on protons at $E_\gamma$ = 9 GeV are given in Table 2 for
light (I), medium (II) and heavy (III) glueball masses. For $a_0(980)$ the total cross section is $59.22$ nb.

\begin{table}[ht]
\begin{center}
\begin{tabular} {|l|c|c|c|}
\hline
Scenario & (I)  & (II)  & (III)  \\
\hline
\hline
 $f_0(1500)$ &  34.98 &  20.25 &  6.61 \\
\hline
$f_0(1710)$ &  0.30 & 0.68 &  5.08 \\
\hline
\end{tabular}
\end{center}
\caption{\it Integrated photoproduction cross sections in nanobarns on protons
at $E_\gamma = 9$ GeV for the three different mixing scenarios:
light (I), medium-weight (II) and heavy glueball (III) (see text).}
\label{tab2}
\end{table}

Concerning the background coming from the decay of vector mesons into two charged mesons it can be get rid by
considering only the all-neutral channels. Namely, the two neutral pions, two neutral $\eta$'s and four neutral
pions decays of the $a_0(980)$, $f_0(1500)$ and $f_0(1710)$. An additional problem is the uncertainty in the
branching fractions of the $f_0(1710)$ \cite{PDG} and its small cross section. On the other hand, the cross
sections for photoproduction of the $f_0(1500)$ on protons are reasonable and its branching fractions are much
better defined. Still about background, the scalars are not produced alone as in the $\pi^0\pi^0$ channel there
is a continuum background arising from the process  of decay of photon in a neutral pion plus a vector meson like
rho or omega with subsequent rescattering of the latter on the proton by $\rho (\omega)$ exchange to give the
second $\pi^0$. This has to be taken into account in a realistic simulation of final state configuration as done
in Ref. \cite{DK}.

In summary, we have studied the photoproduction of the $a_0(980)$, $f_0(1500)$ and $f_0(1710)$ resonances for
photon energies relevant for the GlueX experiment at photon energy of 9 GeV. It would provide novel tests for our
understanding of the nature of the scalar resonances and about current ideas on glueball and $q\bar{q}$ mixing.
The meson differential and integrated cross sections were evaluated and the effect of distinct mixing scenarios
were investigated. Although large backgrounds are expected, the signals could be visible by considering only the
all-neutral channels, that is their decays on $\pi^0\pi^0$, $\eta^0\eta^0$ and $4\pi^0$. The theoretical
uncertainties are still large, with $f_0(1500)$ the more optimistic case. Finally, an experiment in nuclei would
also lead to the $f_0$ and $a_0$ excitation mostly from the collision of protons with protons. The studies in nuclei
would provide information on the meson properties in a nuclear medium, where large modifications are theoretically
expected \cite{MOT}. As a final note on the limitations of the present model, the narrow width approximation for the
scalars could be  insufficient taking into account that the final-state measured particles would be for instance 
$N \pi\pi$, $N K\bar{K}$, $N \eta \pi$ and the energy-dependent width effects would become significant.

\begin{acknowledgments}
 The authors are grateful to A. Donnachie for useful remarks. We also thank Francesco Giacosa and Volker Crede for
 fruitful discussions. This research was supported by CNPq and FAPERGS, Brazil. 
\end{acknowledgments}


\end{document}